\documentclass[12pt,preprint]{aastex}

\usepackage{subfigure}

\shorttitle{}
\shortauthors{Bongiorno et al.}

\begin{document}


\title{A New TeV Binary:  The Discovery of an Orbital Period in HESS~J0632+057}


\author{
S.~D.~Bongiorno\altaffilmark{1,2},
A.~D.~Falcone\altaffilmark{1,2},
M.~Stroh\altaffilmark{1},
J.~Holder\altaffilmark{3},
J.~L.~Skilton\altaffilmark{4},
J.~A.~Hinton\altaffilmark{5},
N.~Gehrels\altaffilmark{6},
J.~Grube\altaffilmark{7}
}

\altaffiltext{1}{Dept. of Astronomy \& Astrophysics, Pennsylvania State Univ., University Park, PA 16802, USA}
\altaffiltext{2}{corresponding author email: sdb210@astro.psu.edu \& afalcone@astro.psu.edu}
\altaffiltext{3}{Dept. of Physics and Astronomy and the Bartol Research Institute, Univ. of Delaware, Newark, DE 19716, USA}
\altaffiltext{4}{Max-Plank-Institut f\"{u}r Kernphysik, Saupferchekweg, Bothe-Labor, D-69117 Heidelberg, Germany}
\altaffiltext{5}{Dept. of Physics \& Astronomy, Univ. of Leicester, Leicester LE1 7RH, UK}
\altaffiltext{6}{NASA Goddard Space Flight Center, Greenbelt, MD, USA}
\altaffiltext{7}{Alder Planetarium, Chicago, IL 60605}


\begin{abstract}
HESS J0632+057 is a variable, point-like source of Very High Energy ($>100$ GeV) gamma-rays located in the Galactic plane.  It is positionally coincident with a Be star, it is a variable radio and X-ray source, has a hard X-ray spectrum, and has low radio flux.  These properties suggest that the object may be a member of the rare class of TeV/X-ray binary systems. The definitive confirmation of this would be the detection of a periodic orbital modulation of the flux at any wavelength. We have obtained {\it Swift} X-ray telescope observations of the source from MJD 54857 to 55647 (Jan. 2009 - Mar. 2011) to test the hypothesis that HESS J0632+057 is an X-ray/TeV binary.  We show that these data exhibit flux modulation with a period of $321 \pm 5$ days and we evaluate the significance of this period by calculating the null hypothesis probability, allowing for stochastic flaring.  This periodicity establishes the binary nature of HESS J0632+057.
\end{abstract}


%% subject header.  Objects should be in the appropriate "individual"
%% headers (e.g. quasars: individual, stars: individual, etc.) with the
%% additional provision that the total number of headers, including each
%% individual object, not exceed six.  The \objectname{} macro, and its
%% alias \object{}, is used to mark each object.  The macro takes the object
%% name as its primary argument.  This name will appear in the paper
%% and serve as the link's anchor in the electronic edition if the name
%% is recognized by the data centers.  The macro also takes an optional
%% argument in parentheses in cases where the data center identification
%% differs from what is to be printed in the paper.



\section{Introduction}
There are currently three confirmed TeV gamma-ray binaries; PSR B1259-63 \citep{aharonian05a}, LS 5039 \citep{aharonian05b}, and LS~I+$61^{\circ}$~303 \citep{albert06, acciari08, acciari09a} as well as a fourth new GeV gamma-ray binary recently detected by {\it Fermi} \citep{corbet11} that could have associated TeV emission.  Evidence at the $4.1 \sigma$ level for Very High Energy (VHE) emission from the stellar mass black hole candidate Cyg X-1 has been reported by \citet{albert07} during a single short flaring episode.  Contained within each of these sources, a young and massive O or B star is orbited by a compact object, either a neutron star or a black hole.  The VHE emission in TeV binary systems is powered either by pulsar winds driving shock acceleration or by mass accretion onto the compact object driving a microquasar jet.  The compact physical size of the emission region leads TeV binaries to be one of the few classes of Galactic VHE sources which appear point-like to TeV instruments.  Following the discovery of several TeV objects with no identified counterparts, HESS J0632+057 was suggested to be a possible member of this exotic class of TeV/X-ray binaries \citep{aharonian07,hinton09}.  We are testing this hypothesis by using X-ray data to search for periodicity from orbital flux modulation.

HESS J0632+057 was first detected as part of the HESS Galactic Plane Survey \citep{aharonian07}.  Located on the edge of the Rosette Nebula within the Monoceros Loop region at R.A. 06h 32' 58.3'', Dec. +05$^{\circ}$ 48' 20'' ($\pm$ 28'' stat., 20'' sys.), the spatial distribution of the source is consistent with that of a point source (rms size $< 2$' at 95\% confidence) \citep{aharonian07}. The source is located on the edge of the 99\% contour of the EGRET unidentified object 3EG J0634+0521, but no object at the position of HESS J0632+057 is listed in the {\it Fermi} LAT point source catalog. MWC 148, a B0pe star, is positionally coincident with the centroid of the HESS position. 

\citet{hinton09} observed the region surrounding HESS J0632+057 on 2007 September 17 with {\it XMM}, obtaining 26 ks of good data.  The observation resulted in the identification of point source XMMU J063259.3+054801, which is positionally coincident with both MWC 148 and HESS J0632+057, and within the 99\% error circle of 1RXS J063258.3+054857.  The source showed significant variability during this short observation ($\approx 7.2$ hours).  During this observation the source showed a mean deabsorbed flux of (5.3 $\pm$ 0.4) $\times 10^{-13}$ erg~cm$^{-2}$ s$^{-1}$ in the 1-10 keV range.  The spectrum was well fit by an absorbed power-law model with a photon index of 1.26$\pm$ 0.04 and a column density of $N_H = 3.1\pm0.3 \times10^{21}$ cm$^{-2}$.  

Beginning 2009 January 26, \citet{falcone10} initiated a campaign to observe XMMU J063259.3+054801 with the {\it Swift} X-ray telescope (XRT; \citet{burrows05}), reporting significant variability in 0.3-10 keV flux. While flux variability was measured on day to month-long timescales, \citet{falcone10} found no evidence for periodicity and concluded that if periodicity was present, it was likely with $P>54$ days.  Additionally, the X-ray source has shown no evidence of short timescale periodic or quasi-periodic variability in the range 0.005 - 800 Hz \citep{rea11}.

During December 2006 - January 2007 and December 2008 - January 2009 \citet{acciari09b} performed follow-up observations of HESS J0632+057 above 1 TeV with VERITAS, yielding flux upper limits well below the values published by HESS.  Together, the VERITAS and HESS observations provide evidence for variability of the gamma-ray flux on time scales of months.  More recently, both VERITAS and MAGIC have detected elevated TeV gamma-ray emission \citep{ong11, mariotti11} during the time period of elevated X-ray flux reported by \citet{falcone11} in February 2011.

We find no report of optical flux modulation in the literature although \citet{aragona10} report temporal variations in the H$\alpha$ emission line profile.  HESS J0632+057 has shown a radio spectral index of $\alpha_r = 0.6 \pm 0.2$ and significant variability at 5 GHz on month-long time scales around a mean flux of 0.3 mJy \citep{skilton09}, however no periodic variability could be detected in these data.  This radio flux is much lower than the typical radio flux expected from a TeV blazar, making such a potential interpretation improbable.  Recently, \citet{moldon11} announced the 1.6 GHz detection of a milliarcsecond scale source coincident with MWC 148.

The point-like nature of the detected TeV source, the excellent positional coincidence with MWC 148 (chance coincidence of $\sim10^{-4}$ according to \citet{aharonian07}), the location on the Galactic plane with a low radio flux, the X-ray binary-like spectral index, and the variable X-ray and gamma-ray emission are all facts that argue in favor of an X-ray binary in association with MWC 148.  Confirmation of HESS J0632+057/MWC 148 as a TeV binary would add a new member to this short list of objects; whereas, refutation of the binary hypothesis would establish an equally interesting class of TeV source. The most direct way to test this binary hypothesis is to search for periodic emission signatures.  In this paper, we report on recent monitoring data taken with {\it Swift}-XRT.

\section{The Observations}
The X-ray Telescope (XRT; \citet{burrows05}) on the {\it{Swift}} observatory \citep{geh04} was used to obtain sensitive observations in the 0.3-10 keV energy band. The typical observation duration was $\sim5$ ks with $\sim1$ week spacing between most observations.  From  16 May 2009 - 15 August 2009 and 10 May 2010 - 21 August 2010 the source was unobservable due to its close proximity to the Sun.  The total {\it{Swift}}-XRT data set includes 463 ks of observations, extending from 2009 January 26 (MJD 54857.1) to 2011 March 27 (MJD 55647.6), representing a baseline of $T=790.5$ days.  All of the observations were obtained in photon counting (PC) mode. All observations beyond MJD 54965 are being reported here for the first time.

\section{Analysis}
We used the most recent versions of the standard {\it Swift} tools and the most recent calibration files available at the time of data processing. In particular, we utilized {\it Swift Software} version 3.7, {\it FTOOLS} version 6.10, and {\it XSPEC} version 12.6.0q. Light curves were generated using {\it xrtgrblc} version 1.5.  Source and background regions, bad CCD detector column correction, and point spread function corrections, and filtering of data was all performed as described in \citet{falcone10}. Since these observations always resulted in XRT count rates $\sim$0.01--0.08 c/s, there was no significant pile-up.

Spectral fitting was done with an absorbed power-law, with the $N_H$ set to the value found by \citet{hinton09} with {\it XMM}. Following the technique of \citet{falcone10}, we calculated rate-to-flux conversion factors for groups of observations, defining the groups such that they contained enough counts to enable good spectral fits (~20 counts/spectral bin) and isolated a single flux state of the source.  Assuming that the observed energy spectrum does not change significantly while the source is in a particular flux state, the rate-to-flux conversion will apply for all observations in each group. The time-dependent flux-rate ratio was then applied to the rate light curve, shown in Figure 1.  Error bars represent the count-rate error scaled by the rate-to-flux conversion factor.

\section{Results}
\label{results}
A spectral fit of the combined data set finds that the data is well fit by an absorbed power-law spectrum with column density set to the {\it XMM} value ($N_H = 3.1\pm0.3 \times10^{21}$ cm$^{-2}$).  This results in an average spectral index of $1.58 \pm 0.06$ and $\chi^2 = 57.9$ with 39 degrees of freedom.  This time averaged spectrum is consistent with the results of \citet{falcone10} and other TeV binaries such as LS~I+$61^{\circ}$~303 \citep{smith09}.  The X-ray light curve, constructed from the entire data set, is shown in Figure \ref{fig:lightcurve_xrt}. Given the observations, we define five different feature types in the variability; low level quiescent emission of $\sim 10^{-12}$ erg cm$^{-2}$ s$^{-1}$, large emission peaks, which are 4-5 times brighter than the quiescent emission, small precursor peaks, which are approximately 2 times brighter than the quiescent emission and precede each large peak, small mid-phase peaks, which are approximately 2.5 times brighter than quiescent emission and appear approximately half way between the large peaks, and dips, which succeed each large peak and represent a decrease in emission by a factor of $\sim 1/3$ below quiescent.  By eye, these emission features appear to be spaced at regular intervals, leading us to believe that their positions reveal an underlying periodicity in the source.  To confirm the presence of periodic modulation, we apply an autocorrelation analysis. The period of the modulation is determined with a peak fitting method, and its statistical significance is established using MC light curve simulations.

\subsection{Autocorrelation}
\label{zdcf}
An autocorrelation analysis is often used as a test for repeating patterns in astronomical time series, as it requires no prior assumptions about the lightcurve profile \citep{ciprini07, xie08}. The Discrete Auto-Correlation Function (DACF), in particular, can be used to study the level of auto-correlation in unevenly sampled datasets without any interpolation or addition of artificial data points \citep{edelson08}.  Figure~\ref{fig:zdcf} shows the result of applying the {\it z}-transformed DACF \citep{alexander97} to the {\it Swift}-XRT dataset. We show both the autocorrelation value, and the autocorrelation divided by its error, which provides an estimate of the significance of the autocorrelation measurement. A clear peak is visible at a time lag of $\sim320$ days, as well as the expected peak at twice this period.

\subsection{Peak Fitting}
\label{peak}
\begin{figure}[ht]
\centering
\subfigure{\label{fig:subfig:flux_lightcurve}
\includegraphics[scale=.8]{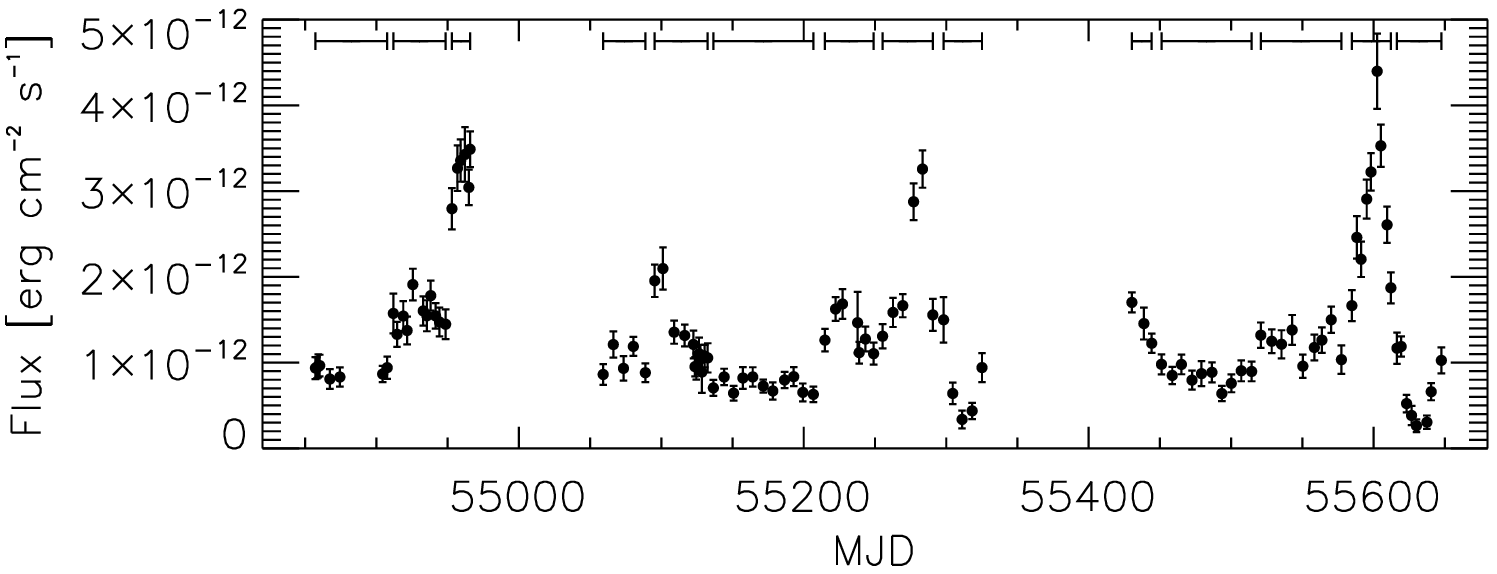}}
\subfigure{\label{fig:subfig:flux_lightcurve_fit}
\includegraphics[scale=.8]{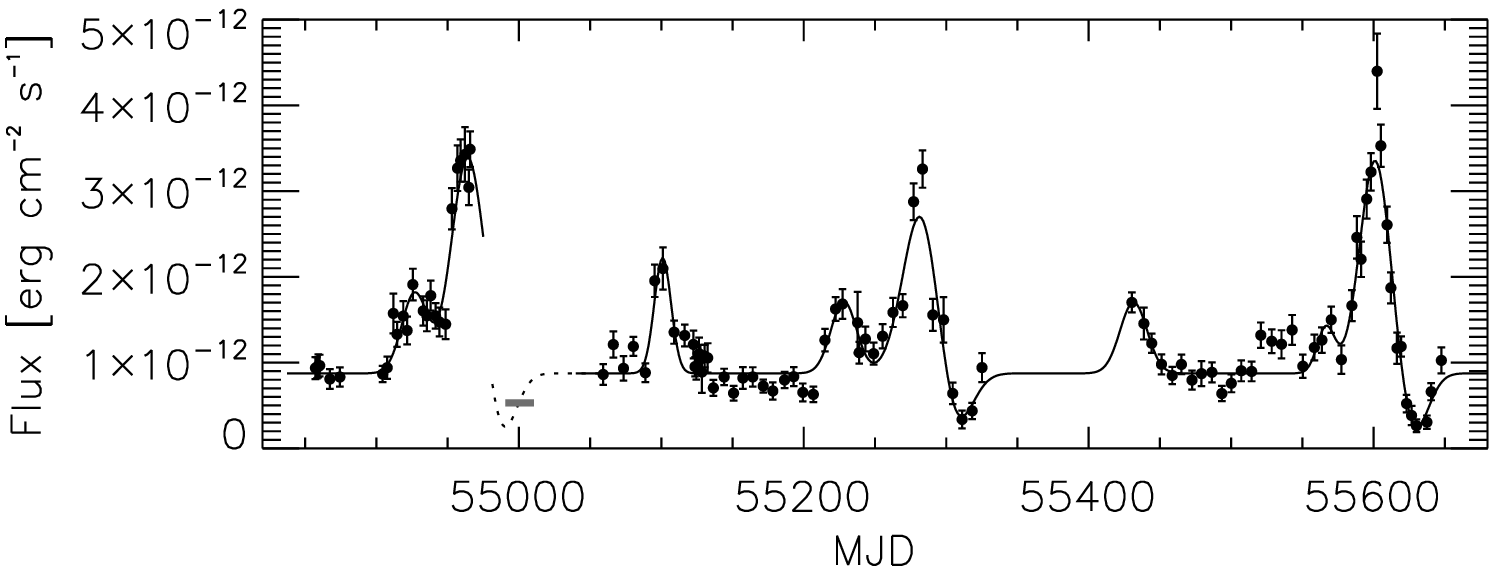}}
\caption{The background-subtracted X-ray light curve of XMMU J063259.3+054801 from the {\it{Swift}}-XRT observations in the 0.3-10 keV band.  Figure \ref{fig:subfig:flux_lightcurve} shows only the data and the time bin boundaries that were used for spectral fitting.  Figure \ref{fig:subfig:flux_lightcurve_fit} shows the same data with the best fit of 10 Gaussians + constant model plotted as a solid line.  Had the source been observable around MJD 55000, we hypothesize that it would have exhibited a dip state.  To show this, we phase the best-fit model in the region of the MJD 55635 dip backwards by two periods and plot it as a dotted line.  We also phase forward the {\it XMM} data taken on 2007 September 17 (MJD 54360) \citep{hinton09} by two periods and plot its error (dominated by period uncertainty) as a gray shaded box.}
\label{fig:lightcurve_xrt}
\end{figure}

\begin{figure}
\centering
\includegraphics[scale=.8]{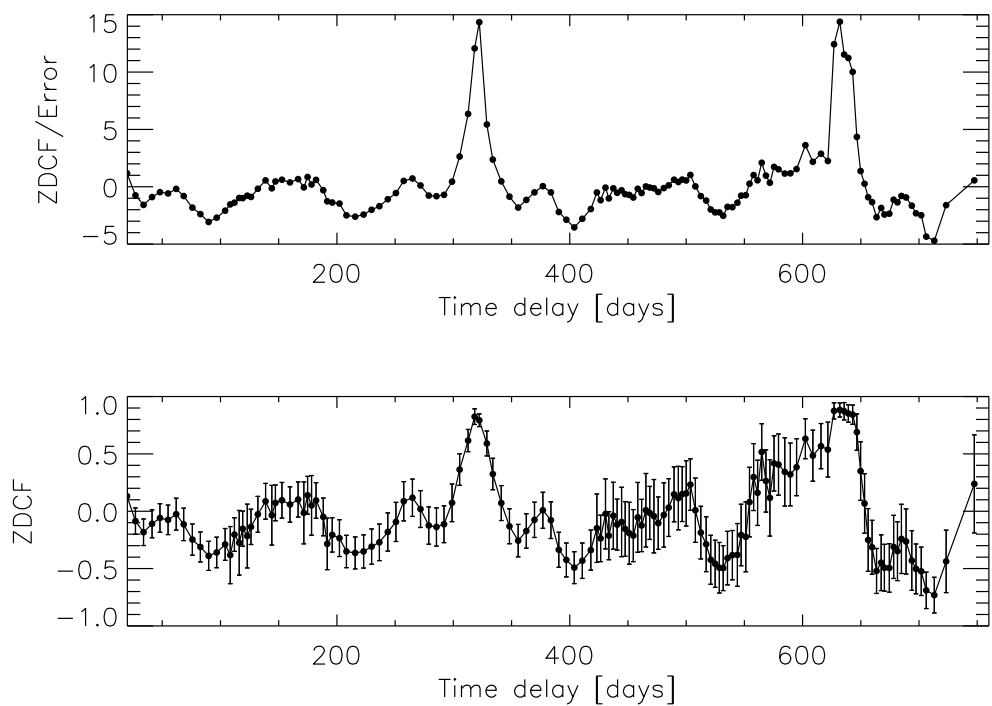}
\caption{The {\it z}-transformed discrete correlation function as a function of time lag, plotted with  error bars (bottom) and divided by error bars (top), to give an estimate of significance.}
\label{fig:zdcf}
\end{figure}

Peak fitting is a simple and robust method for measuring periodicities in data, such as these, where modulation is readily characterized by isolated peaks of enhanced and suppressed emission superimposed upon a quiescent state.  We model the light curve with a sum of 10 Gaussians (one for each peak feature) and a constant (for the quiescent emission).  This results in 31 free parameters.  To obtain the best-fit parameter values, we perform chi-square minimization between the model and data.  The final fit is shown as a solid line in Figure \ref{fig:lightcurve_xrt} and resulted in $\chi^2 = 207.5$ for 79 degrees of freedom.  The large $\chi^2$ is evidence that additional, small-scale variability is present in the light curve and/or that Gaussians do not ideally describe the flares.  While the entire light curve is not ideally fit by this model, it provides a reasonable characterization of the variability and, more importantly, allows us to characterize the timing of the primary features in the light curve.

\begin{figure}
\centering
\includegraphics[scale=0.8]{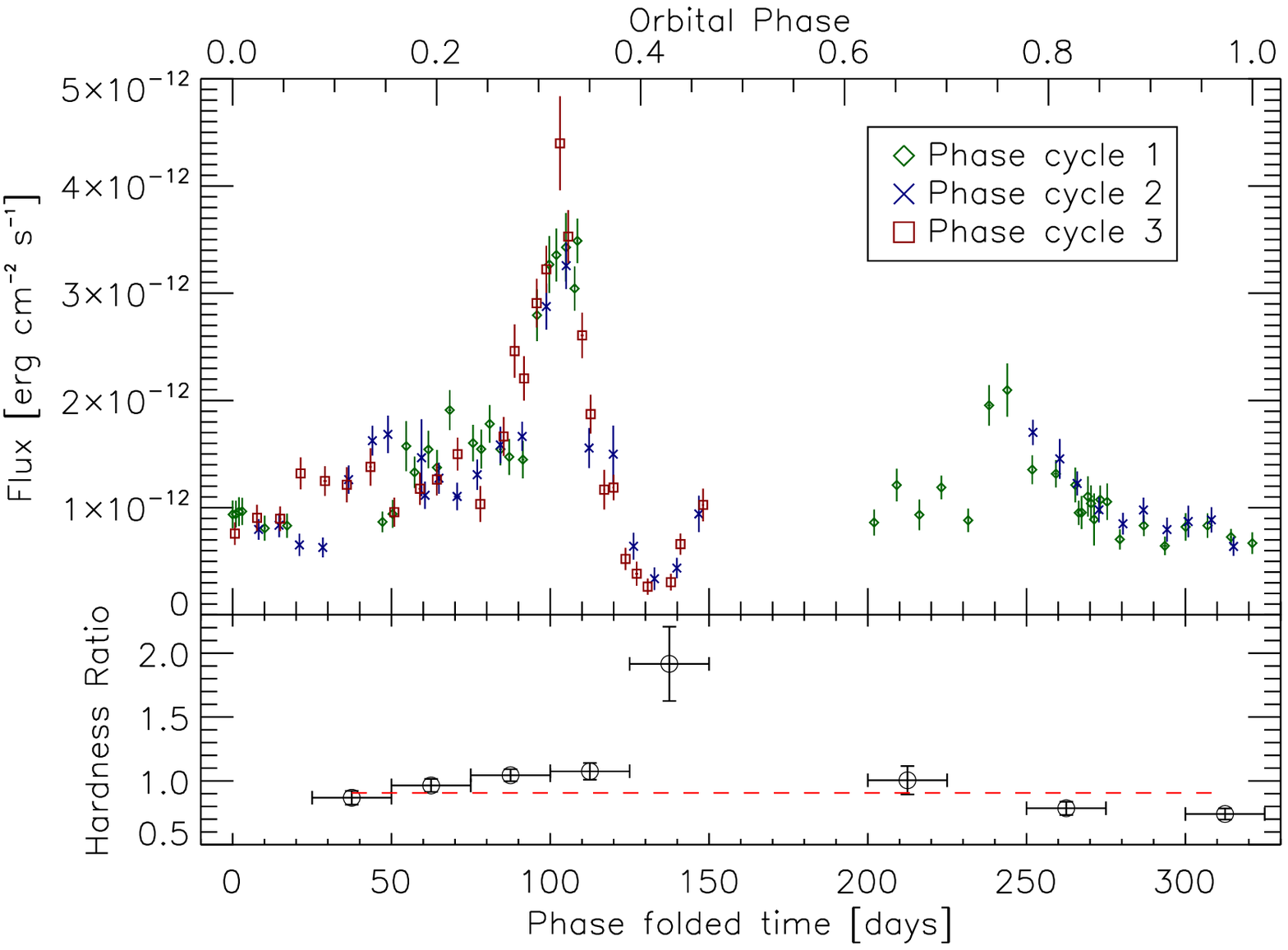}
\caption{The X-ray light curve of XMMU J063259.3+054801 folded over the proposed period of 321 days.  Zero phase has been arbitrarily defined as the date of first observation (MJD 54857).  The three phase cycles that result from this folding are designated with diamond, X, and square symbols, respectively.  The lower panel shows the hardness ratio (2.0-10.0 keV)/(0.3-2.0 keV), folded over the same period and binned at 25 day intervals to improve the signal to noise ratio.  The shown hardness data were fit with a constant (red dashed line), resulting in $\chi^2 = 55.4$ for 7 degrees of freedom, thus confirming variability.}
\label{fig:folded_lightcurve_xrt}
\end{figure}

We propose that the large peaks (MJD 54966.2, 55283.3, and 55601.3) and the dips (MJD 55303.4 and 55630.0) most accurately characterize periodic behavior in the light curve. The small precursor peaks that precede each large peak are unsuitable for measuring the period because they are observed to have variable shape, position (with respect to the large peaks), and height and because their significance above quiescence is small.  The small mid-phase peaks are also unsuitable for measuring the period because the second peak, centered at MJD 55431.8, is poorly sampled, yielding an unreliable peak position estimate.  Averaging the separations between peak centers, using large peaks and dips, we calculate a period of 321 days.  We conservatively estimate the error to be $\pm 5$ days, the approximate average sampling rate during the large peaks and dips.  Using this period, we have plotted the phase-folded light curve in Figure \ref{fig:folded_lightcurve_xrt}.  This shows a strong similarity between modulation observed during different phases, particularly during the large peaks and dips. 

\subsection{Significance of the Period}
\label{significance}
To better estimate the significance of the 321 day periodicity, we performed Monte-Carlo simulations.  By simulating many light curves with flare features similar in size and occurrence frequency to those seen in the observed light curve, we can estimate the probability that the light curve is generated by a source that exhibits the observed flaring pattern every 321 days, and not by a source that flares randomly in time.  To characterize the flaring behavior in the observed light curve, we again fit the 10 Gaussian + constant model to the light curve.  Since we see multiple repetitions of each feature type (large, small, and mid-phase peaks and dips) in the light curve, it is possible to measure not only each feature type's rate of occurrence in the data set, but also estimate the range of values of each type's height and width by using the mean and variance.  We define the rate of feature type occurrence as the number of occurrences per total time where the feature type may have, potentially, been observed.  Given our observation cadence, the only time intervals where peaks or dips could not have been detected were during the date ranges MJD 54970 - 55050 and MJD 55330 - 55424, where the source was too close to the Sun.  During the entire period of {\it Swift} observations when peaks and dips could have been observed, we observed 3 large peaks, 3 small precursor peaks, 2 dips, and 2 small mid-phase peaks.

To generate the appropriate number of features of each type, we sample a random Poisson deviate corresponding to the calculated rate of occurrence.  The shape of each feature is then created with a Gaussian peak of height and width generated by selecting a random deviate from a normal distribution with mean and variance measured in the corresponding feature type in the observed light curve.  The mean position of each peak is generated by selecting a uniformly distributed random deviate on the observation baseline (MJD 54857 to MJD 55647).  The peaks and dips are then combined with the best-fit value for the quiescent emission derived from data.  The final simulated light curve is generated by sampling the model of superposed Gaussian peaks at the bin-center-times when {\it Swift} observed HESS J0632+057.  We then calculate the $\chi^2$ between the simulated light curve and the model that best fit the observed data.  The fraction of simulated light curves with calculated $\chi^2$ less than that of the data represents the probability of generating the observed light curve with a source that is stochastically flaring (with flares having the observed size and rate of occurrence).  In this way, we created $10^9$ simulated light curves with randomly distributed flares and dips, none of which resulted in a $\chi^2$ value less than that of the actual light curve, relative to the fit.  Therefore, we find the false alarm probability for finding this periodic light curve from similar-sized stochastic flaring to be $P< 1\times 10^{-9}$.

\section{Discussion \& Conclusions}
\label{conclusions}
We have detected a $321 \pm 5$ day period in the 0.3-10 keV light curve of the unidentified TeV object HESS J0632+057.  The observed periodicity has been shown to be significant by estimating the chance probability that the observed light curve periodicity is due to stochastic flaring of the source, resulting in a false alarm probability of $P <  1 \times 10^{-9}$.  This implies a binary nature of HESS J0632+057, making it the fourth confirmed TeV binary (there are now 5 confirmed gamma-ray binaries, if one includes the recent announcement by \citet{corbet11} of GeV emission from 1FGL J1018.6-5856).  The $321 \pm 5$ day period makes the system one of the longest period Be star TeV/X-ray binaries and observations by \citet{skilton09} show that HESS J0632+057 may have an unusually low luminosity.

Figure \ref{fig:folded_lightcurve_xrt} shows that the hardness does not vary significantly throughout most of the orbit, but it does reach a significant maximum when the light curve exhibits a dip feature.  A constant line fit of the hardness data results in $\chi^2 = 55.4$ (with 7 dof), which shows that spectral variability is present, particularly during the flux dip. This is consistent with the harder photon index measured by \citet{hinton09} at a time corresponding to a dip feature in the light curve, phased forward by two periods (Figure \ref{fig:lightcurve_xrt}). This hardening of the X-ray spectrum during a flux decrease may be due to increased absorption of the soft X-rays in the source region, or it could be due to an orbital modulation of acceleration site parameters. If increased absorption due to a partial eclipse of the X-ray emission region is the origin of this spectral hardening, then it argues for a geometry of the binary system that allows the Be star and/or its equatorial disk to pass between the X-ray emission region and our line-of-sight. This would imply that the binary orbit is highly inclined and that the emission region is small enough to be eclipsed.  However, this would likely be associated with an increase in N$_H$, which is not observed in these data.  An alternative, and more likely, explanation is that the spectral hardening during the flux dip is unrelated to absorption and instead caused by a change in acceleration site parameters, such as electron injection energies and cooling timescales, as a function of orbital phase.

In addition to the flux dip feature discussed above, the dominant features of the periodic phased X-ray light curve are a large flux peak with a factor of 5-6 flux increase over quiescent flux and a moderate flux peak with a factor of $\sim2.5$ flux increase over quiescent flux.  These peaks are separated in time by about one half period.  The flux dip immediately follows the large flux peak.  Each of these features lasts roughly $\sim4$ weeks.  The recent increase in TeV flux reported by \citet{ong11} and \citet{mariotti11} coincides with the time of the recent large X-ray peak, which could imply related mechanisms such as synchrotron and inverse Compton emission.  By analogy with known X-ray binaries, it is reasonable to assert that the peaks are the result of orbital modulation, but the system geometry and X-ray generation mechanism are not well understood.  If the spectral hardening discussed above is due to absorption from the Be star and the surrounding region, then the geometry would be most easily solved if the large flux peak were due to periastron passage.  However, studies to-date have not found evidence for optical radial velocity shifts in the optical counterpart MWC 148 \citep{aragona10, casares11}.  Additional, radial velocity measurements are required to understand the system's geometry.

While the light curve is not sampled identically in each orbital cycle, making orbit-to-orbit variability difficult to assess, it appears that some variation exists, particularly in the regions preceding each large X-ray flare. This may indicate the presence of other variability timescales related to inhomogeneity in the Be star disk/wind. Other X-ray binaries, e.g. LS~I+$61^{\circ}$~303 \citep{smith09,li11}, show sporadic flaring behavior with a variety of timescales superposed on the binary orbit timescale, and this should be considered as a possible source of orbit-to-orbit variations for HESS J0632+057.  Short timescale flaring analysis provides a way to probe the size of the emission region, as well as the power of the engine that must be feeding the associated acceleration site.  If the dips indeed represent partial eclipses of the compact object by the donor star, this effect may also be used to constrain the size of the emission region and/or the size of a region of X-ray absorption. 

\acknowledgments
This work is supported at Pennsylvania State University by NASA grant NNX10AK92G, contract NAS5-00136, and grant NNX09AU07G.

%% The following command ends your manuscript. LaTeX will ignore any text

\end{document}